\documentclass[a4paper,11pt]{article}
\pdfoutput=1

\usepackage{jcappub} 

\usepackage[T1]{fontenc} 
\usepackage{gensymb}
\usepackage{graphicx}

\usepackage{caption}
\usepackage{subcaption}

\title{Gigaparsec structures are nowhere to be seen in $\Lambda$CDM: an enhanced analysis of LSS in FLAMINGO-10K simulations}

\author[a]{A.M. Lopez,}
\author[a]{R.G. Clowes}

\affiliation[a]{Jeremiah Horrocks Institute,\\University of Lancashire (formerly, Central Lancashire),
Preston, PR1 2HE, United Kingdom}

\emailAdd{amlopez2@uclan.ac.uk}
\emailAdd{rgclowes@uclan.ac.uk}

\abstract{Recently, Sawala et al. 2025 claimed to refute the cosmological significance of the \emph{Giant Arc} based on their analysis of the FLAMINGO-10K simulation data. In our paper here, we highlight several shortcomings of the authors' analysis. We then perform an enhanced analysis on the FLAMINGO-10K simulation data with applications of: the Single-Linkage Hierarchical Clustering (SLHC), the Convex Hull of Member Spheres (CHMS), and the Minimal Spanning Tree (MST) algorithms. Using the full $2.8^3$~Gpc$^3$ FLAMINGO-10K box, with subhaloes at $z=0.7$, and $100$ random realisations (from random subset selections) we find no gigaparsec structures in FLAMINGO-10K, and only a few ultra-large large-scale structures (uLSSs, structures exceeding a maximum pairwise separation of $370$~Mpc). Somewhat surprisingly, we found that the large-scale aspects of the FLAMINGO-10K data could be adequately represented by a Poisson point distribution. The enhanced analysis presented here further supports the remarkable nature of the Giant Arc as a cosmologically-significant structure. Of course, the Giant Arc is also accompanied by a second uLSS, the \emph{Big Ring}. The analysis presented here builds on the work presented by Sawala et al., but amends the application of their statistical assessments. We do not yet know why there appears to be such a large discrepancy between the FLAMINGO-10K data and the observed LSS in Mg~{\sc II} absorbers.
Perhaps the results presented here might suggest that the GA, and especially the GA + BR, presents a more direct challenge to $\Lambda$CDM. In contrast to the conclusion of Sawala et al. that `gigaparsec patterns abound in a $\Lambda$CDM universe' we find that they are nowhere to be seen. }

\begin{document}
\maketitle
\flushbottom

\section{Introduction}

Sawala et al.\ \cite{Sawala2025} have presented an analysis in which they claim that the $\sim 1$~Gpc {\em Giant Arc} discovered by Lopez et al.\ \cite{Lopez2022} is nothing special in cosmological terms, that it is not in tension with the `standard cosmological model' ($\Lambda$CDM), and that `gigaparcsec patterns abound' in a $\Lambda$CDM universe. Based on their simulations, together with random samples, they infer that there is `no reason to believe that the {\em Giant Arc} traces any underlying structure in the Universe'. In this paper we wish to present a somewhat different perspective, in large
part based on our own analysis of their FLAMINGO-10K subhalo data (kindly supplied to us by Till Sawala). We adopt the comoving coordinates of the FLAMINGO-10K data \cite{Sawala2025}, which essentially correspond to the proper coordinates (present epoch) used for the {\em Giant Arc} \cite{Lopez2022}.

We begin by making a few introductory remarks, in part to reinforce some basic points that we have made previously.

(1) The {\em Giant Arc} (GA) \cite{Lopez2022} was discovered {\it visually} in an image of Mg~{\sc II} absorbers (detected in the spectra of background quasars); its central redshift is $z \sim 0.802$. The {\em Big Ring} (BR) \cite{Lopez2024} was similarly discovered {\em visually} in an image of Mg~{\sc II} absorbers; its central redshift is also $z \sim 0.802$, and it lies in the same cosmological neighbourhood as the GA, only $\sim 12^\circ$ north of it.

(2) Analysis was unavoidably {\it post hoc} --- after the event. One of the methods we used (there were others) to describe the structures, statistically and morphologically,  was Single Linkage Hierarchical Clustering / Minimal Spanning Tree (SLHC/MST) or, in the alternative terminology used by Sawala et al. and others, Friends-of-Friends (FoF). SLHC/MST was not involved in the
discovery, which was visual. We emphasised that, in principle, the SLHC/MST approach should be applied only to surveys that have no intrinsic spatial variations. That property of no intrinsic spatial variations of course does not apply strictly to the background quasars so one must proceed cautiously and bear in mind the caveats then associated with the application of SLHC/MST.

(3) We have stated that the GA, and the BR too, add to a growing list of ultra-large large-scale structures (uLSSs) that may challenge the Cosmological Principle, upon which the `standard model' of cosmology, $\Lambda$CDM,  is founded. We emphasise once again that the word `challenge' is not synonymous with `contradict'. It is the growing list that should perhaps be considered further.

(4) If using SLHC/MST/FoF for discovery of candidate structures one must choose the linkage scale appropriately: a useful guide is the mean nearest-neighbour separation for a Poisson distribution, $\bar{r} \approx 0.55 (1/\rho)^{1/3}$. In the case of real observational data, but not simulations, one might then need to make allowances for redshift measurement errors, peculiar velocities, and observational artefacts in the data. For example, a link with an intrinsic separation of 50 Mpc might, because of redshift errors and peculiar velocities, appear at 60 Mpc. Similarly, artefacts such as holes in spatial coverage could cause links to be missed. In contrast, simulations do not have redshift measurement errors or observational artefacts, and although they may incorporate peculiar velocities, the coordinates of points are nevertheless known exactly. Thus, with simulations, only the mean nearest-neighbour separation needs to be considered.

(5) Even with an appropriate choice of linkage scale, the SLHC/MST/FoF finds {\em candidates} for structures. One might then select, via some algorithmic method for example, the candidates that appear most plausible and meriting further investigation. Even then, with plausible candidates, a sensible approach is to seek independent corroboration if suitable datasets are available. For example, with the GA  we found that the distribution of quasars in the same redshift slice provided further support. In the case of the BR, further support was provided by both quasars and DESI clusters. (But note that quasars, and especially DESI clusters, have larger redshift uncertainties than the Mg~{\sc II} absorbers and so will be blurrier tracers of structures.)

(6) The standard model of cosmology, $\Lambda$CDM, is more accurately described as the concordance model --- because it fits a wide range of observations, not because it is understood. It is based on the assumption of the Cosmological Principle and the three hypotheses of inflation, dark energy, and cold dark matter. Perhaps there are important features of the Universe that are not incorporated in $\Lambda$CDM. In allowing for that possibility, it makes sense to investigate apparently anomalous features such as the GA and BR in case they should happen to point the way. 

While the GA and BR might in due course be found to be perfectly compatible with $\Lambda$CDM, the analysis of Sawala et al. \cite{Sawala2025} does not, in fact, succeed in making that finding. We elaborate below.

\section{Minimal Spanning Tree methods}
\label{sec:MST}
In cosmology, Minimal Spanning Tree (MST) methods such as the Single-Linkage Hierarchical Clustering (SLHC) algorithm, or, as often otherwise known, the Friends-of-Friends (FoF) algorithm, are useful statistical tools for identifying candidates for large-scale structures (LSSs).
Each method requires specifying a linkage scale, 
which should be appropriate (physically sensible) to the data.
Clearly, the number of candidates, and the membership/size of each candidate, will be sensitive to the choice of linkage scale.

\subsection{Linkage scales}
\label{subsec:linkage_scales}
The mean nearest-neighbour separation (for a Poisson distribution) is $ \bar{r}  \approx 0.55 (1/\rho)^{1/3}$, for $3D$ point data with a field number density $\rho$~\footnote{Note that Sawala et al. instead consider their `mean interparticle distance' as $(1/\rho)^{1/3}$ (as for points on a regular lattice).}.
For real (not simulated) data, other factors should be considered too, such as redshift errors and artefacts in the data.
When applying the SLHC algorithm to the SDSS DR7QSO quasars for the analysis of the two large quasar groups (LQG): the Huge-LQG \cite{Clowes2013} and Clowes-Campusano LQG \cite{Clowes2012} --- the linkage scale was increased from the mean nearest-neighbour separation to allow for the redshift errors and for peculiar velocities.
Similarly, for the two Mg~{\sc II} structures: the Giant Arc \cite{Lopez2022} and the Big Ring \cite{Lopez2024} --- the chosen linkage scale was increased to allow for the gaps in the data from the varying availability of background probes (quasars), as well as errors from peculiar velocities (the redshift errors being relatively small).
We show here why the chosen linkage scale of $95$~Mpc is appropriate for the Mg~{\sc II} data, but is \emph{not} appropriate for simulated data at the same field density.

For the Giant Arc field that was investigated by Sawala et al.\ \cite{Sawala2025}, $V= (1541 \times 1615 \times 338)$~Mpc$^3$ 
$= 841 \times 10^{6}$~Mpc$^{3}$, $N=504$, and $\rho = N/V = 5.99 \times 10^{-7}$~Mpc$^{-3}$.
The FLAMINGO-10K data, at $z=0.7$, for subhaloes in the mass range $1$--$5 \times 10^{12}\,$M$_{\odot}$, has a number density $\sim 2000$ times that of the Mg~{\sc II} GA field.
The authors then randomly select subsets of subhaloes to replicate the number density of the GA field. 
For the FLAMINGO-10K data, 
where the location and redshifts are exactly known,
a linkage scale $\bar{r}  \sim 65$~Mpc would be appropriate when applying the FoF (or SLHC) algorithm.
That is, additional considerations to account for gaps in the data,  redshift errors, and peculiar velocities are not necessary here. 

Consider a Mg~{\sc II} field with the same field density.
A Mg~{\sc II} absorber can be detected only where there is an available background probe.
A low-density region of background probes will likely lead to a low-density region of Mg~{\sc II} absorbers.
When analysing Mg~{\sc II} fields, we aim for an approximately uniform selection of background probes by making appropriate $i$-magnitude cuts, $S/N$ cuts, and/or avoiding harsh SDSS-density borders \cite{Lopez2022, Lopez2024}.
However, the inhomogeneities of the background probes still require careful consideration when analysing the Mg~{\sc II} density maps.
In a particularly low-density region of probes, where the connectivity of Mg~{\sc II} would be under-represented, the linkage scale would need to jump across the low-density patches to connect one data point to its nearest neighbour.

\begin{figure}
    \centering
    \includegraphics[width=0.8\linewidth]{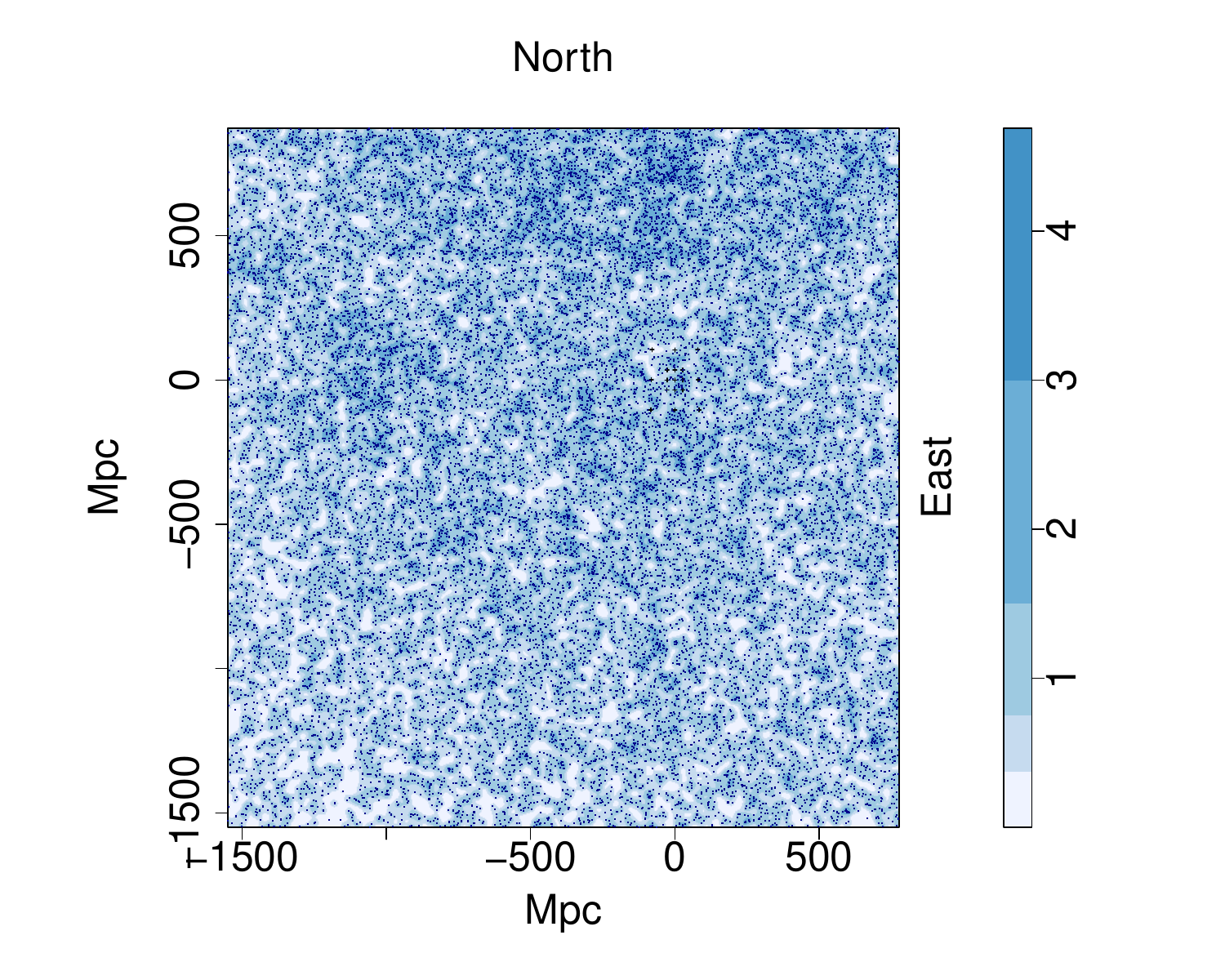}
    \caption{The tangent-plane distribution of probes (background quasars) corresponding to the GA large FOV
with the redshift condition z > 0.862 — i.e., the probes that could in principle show Mg II
absorbers in the GA field. The axes are in proper size, present epoch Mpc, scaled according to the central redshift of the GA field (i.e., $z=0.802$). The blue contours, increasing by a factor of two, represent
the density distribution of the probes which have been smoothed using a Gaussian kernel of
$\sigma = 11$~Mpc. The small, black crosses at $x, y = 0, 0$ roughly correspond to the tangent-plane $x$--$y$ position of the GA. }
    \label{fig:GA_probes_big_FOV}
\end{figure}

To calculate this effect, we count the number of probes in cells of size $65 \times 65$~Mpc$^2$ in the $2D$ distribution of background probes that correspond to the GA Mg~{\sc II} field and measure the variation. 
Specifically, we are looking at the background probes that correspond to the \emph{larger} field of view (FOV) containing the GA (Figure \ref{fig:GA_probes_big_FOV}, or Figure $4$ of the GA paper \cite{Lopez2022}).

The mean cell-count of the probes is $15.6$ and the standard deviation is $5.74$. 
If we reasonably consider the ratio of cell-counts that are underdense that lie $1.5\sigma$ from the mean, then $\bar{n} / (\bar{n} - 1.5 \sigma) = 2.23$.
That is, there is a factor of $2.23$ between an average-density patch of probes in the field and a plausible low-density patch of probes.
The $2.23$ factor then scales with density, so that the adjusted linkage scale, to account for gaps in the data from the inhomogeneous background probes, is $65.2$~Mpc~$\times (2.23)^{1/3}=85.2$.

Next, consider the redshift errors and peculiar velocities of the Mg~{\sc II} absorbers.
Plausible values of peculiar velocities might be $\sim 400$~kms$^{-1}$, corresponding to a redshift difference of $\Delta z= 0.0024$.
Redshift errors for the Mg~{\sc II} absorbers, on the other hand, are considerably smaller than this and on the order of $10^{-4}$--$10^{-5}$, so we may ignore the effects of redshift errors on linkage scale. 
Considering just the peculiar velocities, a redshift difference $\Delta z= 0.0024$ at $z \sim 0.8$ corresponds to a physical $z$ error of $\sim 7$~Mpc. For pairwise separations, combining two such errors in quadrature gives $\sim 9.9$~Mpc. The final, adjusted linkage scale then becomes $\sim (85.2 + 9.9)$~Mpc~$=95.1$~Mpc, or $\sim 95$~Mpc.

The linkage scale that is appropriate for the Mg~{\sc II} data will not be appropriate for simulated data at the same field density. 
For the analysis in this paper we shall use an appropriate linkage scale of $65$~Mpc when analysing the FLAMINGO-10K data.

\subsection{Seeing patterns in noise --- simulations}
Whether investigating real or simulated data, the process of analysing LSS is \emph{fundamentally incomplete} at the stage of having identified \emph{candidate} structures with some use of MST/SLHC/FoF --- further application of a reality-assessment test, or support from independent data sources, is necessary, otherwise the candidate structures are meaningless (one might simply have detected patterns in noise).
Since Sawala et al. do not investigate further any of their candidate structures or `GA-analogues', they have not shown what cosmological relevance, if any, their detected patterns might have.

An important, but unfortunately commonly-missed, aspect in LSS comparisons with random / simulated data \cite{Nadathur2013, Park2012, Park2015} is the independent tests of the candidate structures (or even the candidate fields containing the candidate structures).
This point has been made previously by Marinello et al. \cite{Marinello2016} --- these authors analysed both the significance, and the extreme value analysis, in mock large quasar groups (LQG) in the Horizon Run 2 simulation data and demonstrated that the Nadathur \cite{Nadathur2013} approach (followed by Sawala et al.) was insufficient for determining the significance of the Huge-LQG.

\section{The Sawala et al. analysis}
Sawala et al. 2025 \cite{Sawala2025} claim to have refuted the cosmological significance of the GA by showing that gigaparsec patterns are easily reproduced in the new FLAMINGO-10K simulations. 
They make this claim by attempting to follow one of the statistical methods that were presented in the GA discovery paper.
However, there are some procedures in the Sawala et al. analysis that make their comparison of the FLAMINGO-10K simulated data with the Mg~{\sc II} data inappropriate.

In Section~\ref{subsec:linkage_scales} above, we discussed why the linkage scale that was applied to the Mg~{\sc II} data ($95$~Mpc) is not appropriate for the simulated data at the same field density ($65$~Mpc is appropriate). 
Next, we highlight several other shortcomings of the Sawala et al. analysis.

\subsection{The GA `analogues'}
In their paper, Sawala et al. report the many `analogues' to the GA (specifically, GA-main, see original discovery paper \cite{Lopez2022}) that they claim to have identified in the FLAMINGO-$10$K simulations. They say that such analogues `abound'.

In the eight example, claimed analogues of the GA that they show in their Figure~1, none has \emph{both}  membership and number-overdensity that exceed those of the GA.
There are two that exceed the GA membership, and one that has greater number-overdensity (from the CHMS calculation; see Section~\ref{sec:CHMS_vs_MST}) to the GA.
The authors claim that there are GA-analogues in almost every sample of the FLAMINGO-10K simulations, but fail to present one example `GA-analogue' that truly does resemble or exceed the real GA in size, membership and overdensity.
Moreover, since the authors do not assess their candidate structures with some independent reality-assessment test, it remains uncertain whether any of the `GA-analogues' represent significant deviations from random expectations as the real GA does --- that is, we cannot know whether their `GA-analogues' are cosmologically interesting. 
The authors appear to demonstrate the difficulty in reproducing a structure equal to or greater than the GA in both membership and number overdensity, which would actually appear to support the remarkable nature of the GA.
(Note that, since the authors use an inappropriate linkage scale, their candidate structures favour chaining \cite{MurtaghHeck}.
Therefore, even if Sawala were able to reproduce many GA-analogues in their analysis, the comparison of these `structures' with the real GA would be inappropriate, since their `candidates' would not be sound candidate structures  --- see Section~\ref{sec:chaining} for more details.)

Next, consider the cumulative volumes of simulated FLAMINGO-10K data analysed by the authors compared with the volume, at the time of discovery, of Mg~{\sc II} data that was analysed for LSS.
Sawala et al. begin with the standard FLAMINGO-10K box of size $2.8^3$~Gpc$^3$.
They then select thin slices of size $2.8 \times 2.8 \times 0.338$~Gpc$^3$ from each of the three coordinate axes.
The publicly available code\footnote{https://github.com/TillSawala/GiantArc} suggests that the authors then select $1000$ overlapping slices along each of the three coordinate axes in the first random sampling to produce their `GA-analogues' in their Figure~$1$.
The volume of the RA-Dec-$z$ box of the GA large FOV investigated by Lopez et al. is roughly the same as the Sawala et al. slices, so they investigate roughly $3000$ times as many volumes as were investigated by Lopez et al. in the GA analysis. 

Then, for their structure persistence (see below), the authors investigate $3 \times 7$ \emph{non}-overlapping redshift slices (seven slices per coordinate axis). From the publicly available code, we deduce that the authors then repeat the random subhalo selections $1000$ times. 
That would be $21 000$ times as many volumes as the GA big FOV that we investigated.

With such a large look-elsewhere effect in their FLAMINGO-10K analyses, Sawala et al. presumably expected to easily find examples of candidate structures that simultaneously exceeded the membership and number-overdensity of the observed GA. 
However, in practice, they appear to have been unsuccessful in reproducing even one GA counterpart. Incidentally, and hardly surprisingly (since the chance of doing so is plausibly infinitesimal), they appear
to have been unsuccessful also in simultaneously reproducing the Big Ring that lies in the exact same redshift slice and field of view as the GA.

\subsection{A note on structure persistence}
\label{sec:structure_persistence}
Sawala et al. attempt to further assess the reality of their GA-analogues by considering structure persistence.
Since they randomly select a subset of subhaloes from the full set of subhaloes at $z=0.7$ in the $2.8^3$~Gpc$^3$ FLAMINGO-10K cube, they can check for structures persisting across the random selections as an indication of real underlying structure.
They do not find structure persistence in their analysis, so they conclude that their GA-analogues are not real candidates, and just noise.
Note that finding noise in simulations is not a particularly remarkable discovery. 

Sawala et al. then incorrectly extend their conclusion to mean that the GA is not then tracing any real underlying structure.
Consider what appears to be their argument in reaching this false conclusion: (i) structure persistence is an indication of real underlying structure; (ii) there is no structure persistence in the FLAMINGO-10K simulations; (iii) \emph{therefore} the GA must not be tracing any real underlying structure.
Points (i) and (ii) are true.
However, clearly, points (i) and (ii) do \emph{not} lead to point (iii).
The authors have \emph{not} shown that the GA is not persisting:
they have shown that \emph{their} simulation `structures' are not persisting. In other words, they have simply shown that you can find patterns in noise in FLAMINGO-10K simulations.
Their non-sequitur (iii) has no bearing on the real GA.

\subsection{Subtleties of redshift slices}

Sawala et al. suggest that the GA field is inherently highly anisotropic because the box dimensions have much larger RA-Dec (or tangent-plane $x$-$y$) dimensions than the redshift dimensions, and because the linkage scale ($95$~Mpc) is too close to the shortest dimension ($338$~Mpc) of the redshift slice.
However, the nature of the observational data means that we are restricted to analysing small redshift slices.
What Sawala et al. fail to recognise is that their simulated data are not affected by redshift evolution  (all of their simulated data are conveniently acquired at one redshift only, $z=0.7$), and so there is then no need to restrict their box dimensions to replicate a thin redshift slice.
In fact, with their choice of an inappropriately large linkage scale
the authors' analysis  will be subject to the undesirable effects of chaining (Section~\ref{sec:chaining}), and together with their unnecessarily-thin box slices, their chained structures will inevitably and unavoidably be anisotropic. Their  `candidate structures' are thus unlikely to be of any relevance to the GA observations or to the real Universe in general.

Consider the redshift evolution, from the near side to the far side of a Mg~{\sc II} redshift slice, corresponding to a box that spans (proper size, present epoch) $2.8^3$~Gpc$^3$ (the dimensions of the FLAMINGO-10K box). At the central redshift $z=0.802$ of the GA, the near side of the box would be at $z\sim 0.3$ and the far side would be at $z\sim 1.3$.
Clearly, the evolution of structure with redshift in such a large redshift interval would become very important and very complicated.

The Mg~{\sc II} absorbers are observed in background quasars. 
That is, when looking at a redshift slice containing Mg~{\sc II} absorbers, we have only the quasars that are further than the far side of the Mg~{\sc II} redshift slice as probes of the Mg~{\sc II} absorbers.  In a redshift slice that is too large, we would be unnecessarily restricting the availability of background probes and the ability to detect the intervening matter.

The redshift evolution in redshift slices and the dependence of Mg~{\sc II} absorbers on the available background probes (plus measurement uncertainties of course) are unavoidable, observational and practical considerations of working with the (real) Mg~{\sc II} data. 
With idealised, simulated data, however, Sawala et al. have no such considerations and, instead of following our analysis quite so closely, they could perhaps more productively have adapted theirs to that fact.

In Section~\ref{sec:results} we do not restrict the box dimensions containing the FLAMINGO-10K subhaloes to mimic the dimensions of the Mg~{\sc II} redshift slice. 
Applying the statistical tests to the random samples of FLAMINGO-10K subhaloes in the entire $2.8^3$~Gpc$^3$ box will allow for larger candidate structures to be identified (we are searching for candidate structures with both greater membership and greater overdensity than that of the GA-main).

\subsection{CHMS versus MST --- significance and overdensities}
\label{sec:CHMS_vs_MST}

The Convex Hull of Member Spheres (CHMS) method \cite{Clowes2012} is intended to assess the significance of candidate structures. It works by first drawing a sphere of fixed radius around each of the $N$ candidate points; the radius of the sphere is set to half of the mean MST edge-separation of the $N$ points. Then the convex hull of the $N$ member spheres is constructed. It thus contains all of the member points and their associated spheres in a uniquely-specified volume. Similarly, $1000$ CHMS volumes are constructed of $N$ points randomly distributed in a cube, at a field density equal to the control-field density. The distribution of the $1000$ CHMS volumes allows the assessment of the probability of obtaining randomly a CHMS volume equal to or smaller than that observed.

A shortcoming of the CHMS method is that the unique volumes encompassing candidate structures can often incorporate very large, empty volumes.
For example, the volume containing points in an arc will include the volume of the empty space encompassed by the arc.
Additionally, if an inappropriately-large linkage scale is applied to a set of points, the SLHC will lead to chaining \cite{MurtaghHeck}, and the CHMS volumes of these huge `candidate structures' will be excessively large and contain several empty volumes between the points, especially between the chained groups.
The huge `candidate structures' might easily have large (CHMS) volume \emph{under}densities --- intuitively not real ---, and the CHMS assessment of significance becomes meaningless.
A \emph{sound} candidate structure, by the nature of detection via an \emph{appropriate} linkage scale, will always be overdense. 

The SLHC method is intended to find clumps and clustered points in the data. Curvature can lead to underestimates of significance by the CHMS method, but an inappropriate linkage scale followed by chaining can invalidate the method entirely. The CHMS method is best applied to small, clumpy structures, rather than excessively large, or curved, filamentary structures. 

For strongly curved or filamentary structures, the MST method \cite{Pilipenko2007} for estimating overdensities and significances might be more appropriate.
In our original discovery paper, we gave preference to the MST method, as we found it was more appropriate for dealing with the curved, filamentary morphology of the Giant Arc. 
Sawala et al., however, concentrate on only the CHMS method. 
With this MST method, it is again necessary to apply an appropriate linkage scale to the data points to avoid chaining (Section~\ref{sec:chaining}).  

\subsubsection{Number overdensity and matter overdensity}
We add a short note here on number overdensity and matter overdensity.
As Sawala et al. point out, the number overdensity is often larger than the matter overdensity due to bias.
Sawala et al. use the simulation particle data to compute the matter overdensity, but it is not entirely clear what bias factor this results in.
The bias factor will be different for different tracers of the matter.
For example, bright, luminous objects are usually considered very biased tracers of matter.
For the GA (and other Mg~{\sc II} LSSs), two caveats that may need to be considered are: specifically the bias factor of the Mg~{\sc II} absorbers, and the large scales on which we are observing the matter.

Several studies have investigated the bias factor in Mg~{\sc II} and have shown that Mg~{\sc II} can overestimate the matter overdensity from anywhere between $10 \%$ and up to a factor of $2$, resembling the bias factor seen in massive galaxies (CMASS galaxies from the SDSS) \cite{Gauthier2009, Lundgren2009, Perez-Rofols2015}. 
In particular, there appears to be a slight anti-correlation in the biasing factor based on the strength (equivalent width) of the absorbers \cite{Gauthier2009, Lundgren2009}.
(Note that, in our Mg~{\sc II} LSSs, our EWs can range from $W_r < 0.3$ to $W_r > 1$.)
However, more importantly, it appears that these studies focus on the biasing factor on small scales (relative to uLSSs at $10^2$--$10^3$~Mpc) up to $\sim 30$~Mpc. 
As we approach very large scales biasing factors should become smaller. 
Guo and Jing 2009 \cite{Guo2009} showed that at large scales, $\sim 90$~Mpc, the biasing factor becomes constant and close to $1$ (see their Figure~7).
Given that we are concerned only with matter overdensity on very large scales, the biasing factor should not be a particular concern.
Further studies on the biasing factor in Mg~{\sc II} absorbers on the scales of LSSs and uLSSs might, however, be useful, to test for any putative discrepancy between number overdensity and matter overdensity. 

Finally, note that since the patterns identified by Sawala et al. favour chaining (see below), matter overdensities close to $0$, or even matter underdensities, would be expected.

\subsection{Chaining in SLHCs}
\label{sec:chaining}
Chaining \cite{MurtaghHeck} in SLHCs is an undesirable effect of artificial clustering after applying the SLHC with an inappropriate linkage scale to point data.

In the context of the work presented here, suppose that there are some potentially real clumps at a linkage scale of 65 Mpc. 
By using an inappropriate linkage scale of 95 Mpc, the real, \emph{separate} clumps and other points could be chained together. 
Since Sawala et al. are favouring chaining with their inappropriately-large linkage scale, their gigaparsec patterns then abound, but more importantly, they appear to have no cosmological import whatsoever. 

In our original discovery paper of the Giant Arc, we discussed what results from applying linkage scales at (85, 90, 95, 100 and 105)~Mpc. 
At the smallest linkage scale, only four candidate structures were identified, each having very small memberships. 
Even so, the largest candidate structure was that which belonged to the GA-main, and it was also the only candidate structure in the field identified with a CHMS-significance $> 3\sigma$.
At the largest linkage scale, only three candidate structures were identified, with the largest structure containing $133$ absorber members (including the members of the GA).
Clearly, this is not a viable candidate structure, and we would not treat it as such; with this inappropriately-large linkage scale we have simply, and meaninglessly, chained together separate clumps.
At the remaining linkage scales, the GA-main was identified as a single candidate structure containing the majority of the GA-main member points. Increasing the linkage scale might allow for the inclusion of additional absorbers, which is reasonable considering the errors in the real data from peculiar velocities, and non-uniformities in background quasars (gaps in the data), but it does not incorporate the chaining of groups until we reach an inappropriately-large linkage scale (e.g., $105$~Mpc). 
Again, for real data, this upper limit will be larger than it is for simulated data, which needs to account only for the field density.

Hypothetically, if the assessment of the GA from applying SLHC / CHMS / MST was a result of chaining, one would observe separate, distinct clumps added in with the `candidate structure' upon increasing the linkage scale (or, in reverse, one would observe separate clumps removed upon decreasing the linkage scale). 
That is, chaining will result in separate clumps being joined together, whereas, without chaining, a true candidate structure should remain roughly the same (visually, and in terms of its member points).
Of course, since the SLHC method is not strictly appropriate for the Mg~{\sc II} data, we can use other checks of validity such as independent corroboration and additional statistical tests.

\subsection{Representing Mg~{\sc II} gas with simple CDM particles and subhaloes}

Mg~{\sc II} gas tracing cosmological large-scale structure is complex.
It is not yet well-understood how the gas dynamics affect the large-scale distribution of the Mg~{\sc II} gas.
For example, the Mg~{\sc II} is well known to trace galaxies and galaxy clusters, but the gas may be subject to high velocities that can move this gas beyond the host-galaxy dark matter halo \cite{Kauffman2017}.
This study \cite{Kauffman2017}, comparing the observational large-scale structure traced by Mg~{\sc II} gas with predictions from cosmological simulations, suggests that Mg~{\sc II} cannot be modelled by dark matter particles or subhaloes.

It is worth noting that Sawala et al. use the {\sc HBT+} algorithm \cite{Han2018} in FLAMINGO-10K to identify the subhaloes that they use for their analysis.
However, since then, improvements have been made to {\sc HBT+}, and a new version has been introduced: {\sc HBT-HERONS} \cite{Moreno2025}.
The new version {\sc HBT-HERONS} was designed to improve the tracking of subhaloes.
In the previous version, HBT+, the tracer particles were not restricted to any particular matter tracer, which meant that tracer particles attached to gas or blackholes could be lost owing to ejection outflows or blackhole mergers, which can lead to the sudden loss of well-resolved substructure \cite{Moreno2025}.
Perhaps this loss of substructure would be reflected in the LSS too.

In our analysis below, we do not find any gigaparsec structures, and find only a few uLSSs;
we suggest that these FLAMINGO-10K subhaloes could be under-representing what might be expected in an actual $\Lambda$CDM universe.
A reason for this under-representation could be the particular subhalo-finder being used, as discussed above.
However, ultimately, simulations are complex (and so is the real Universe), so conceivably there are ways in which improvements could be made that would eventually result in an agreement of LSS analysis in simulations with observations.

\section{Reanalysing the FLAMINGO-10K simulations}
\label{sec:results}

Here we apply our sequence of the SLHC algorithm and the CHMS/MST significance and overdensity calculations to the same FLAMINGO-10K subhalo data, kindly provided by Till Sawala.
The results of the significances, overdensities, memberships and maximum pairwise separations of identified candidate structures are shown in Figures~\ref{fig:CHMS_MST_signif_65Mpc}--\ref{fig:max_pairwise}.

\begin{figure}
    \centering
    \includegraphics[width=\linewidth]{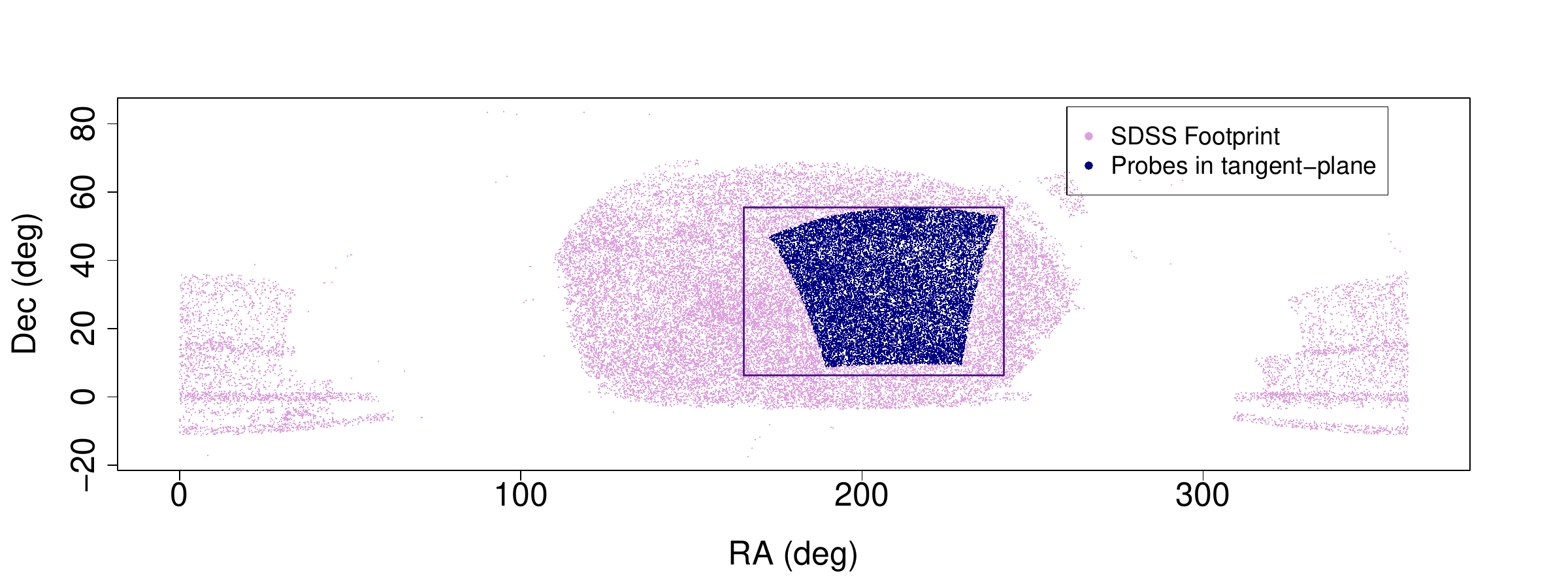}
    \caption{The SDSS DR7QSO and DR12Q combined footprint shown in lilac points. The dark-blue points correspond to the tangent-plane distribution of probes (background quasars) in the GA \emph{larger} field of view (Figure~\ref{fig:GA_probes_big_FOV}). The dark-purple border indicates the original RA-Dec selection, from which the tangent-plane points are drawn. The curvature in the dark-blue points shows the tangent-plane warping effects (note, this is not a problem for simulated box data). In the SLHC/CHMS tests, we use the RA-Dec-$z$ selection of points, i.e., the points inside the dark-purple border.  }
    \label{fig:GA_big_FOV_in_SDSS}
\end{figure}

\begin{figure}
    \centering
    \includegraphics[width=0.8\linewidth]{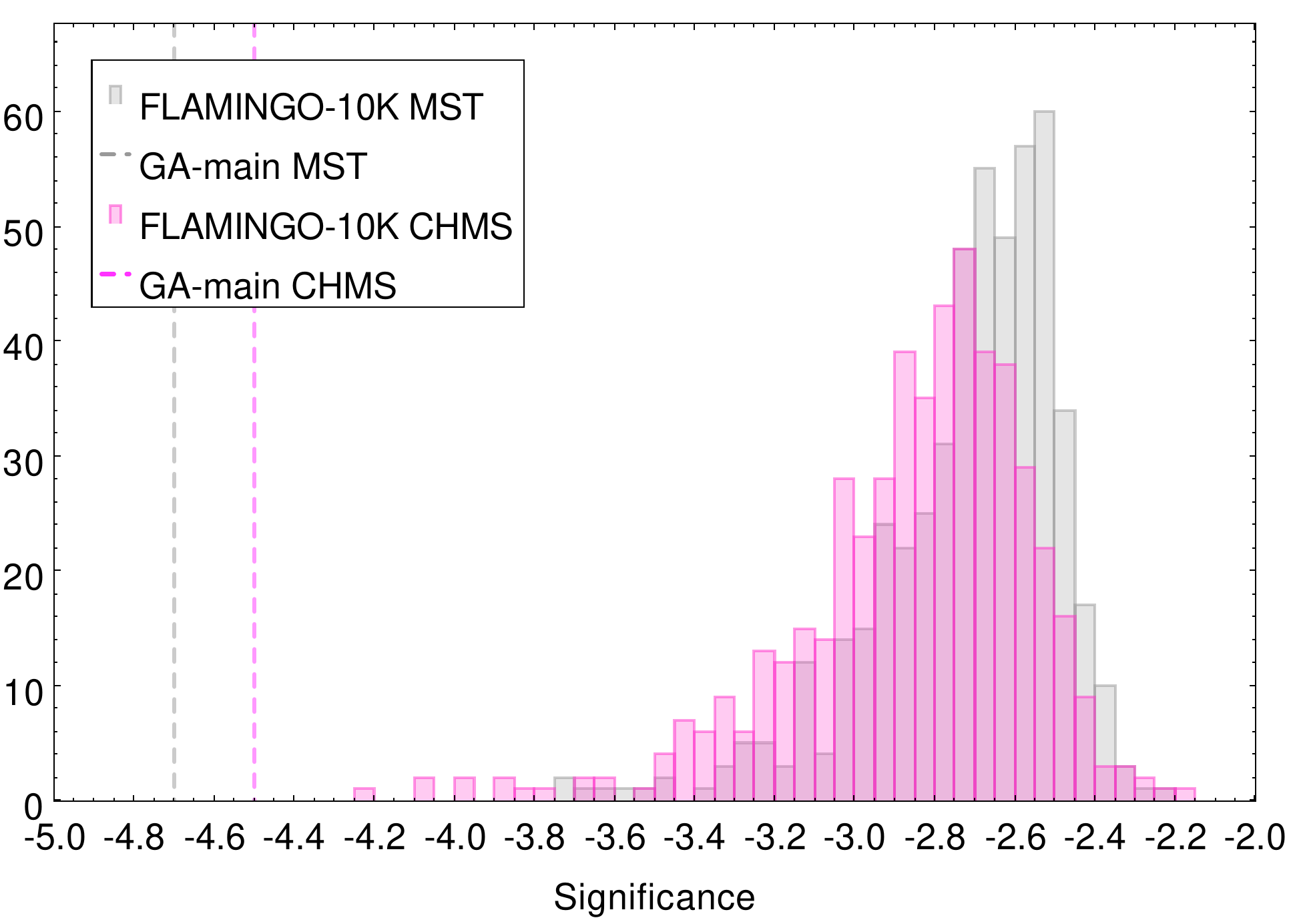}
    \caption{Histogram of the MST and CHMS significances (in grey and pink respectively) in $100$ cubes ($2800^3$ Mpc$^3$) of randomly-selected subhaloes in the FLAMINGO-10K simulation data. We applied a linkage scale of $65$~Mpc in each of the $100$ cubes, which is an appropriate choice for simulated data at this field density (see main text). We specified membership $N \ge 10$. The histograms have similar profiles and are closely aligned, indicating that the results of the MST and CHMS significances are in agreement. The GA-main MST and CHMS significances are labelled at $-4.7 \sigma$ and $-4.5 \sigma$ with grey and pink vertical, dashed lines, respectively. (Our convention is that significances are negative for overdensities.) The GA-main MST and CHMS significances are both significant departures 
    from expectations with simulated FLAMINGO-10K data. }
    \label{fig:CHMS_MST_signif_65Mpc}
\end{figure}

\begin{figure}
    \centering
    \includegraphics[width=0.8\linewidth]{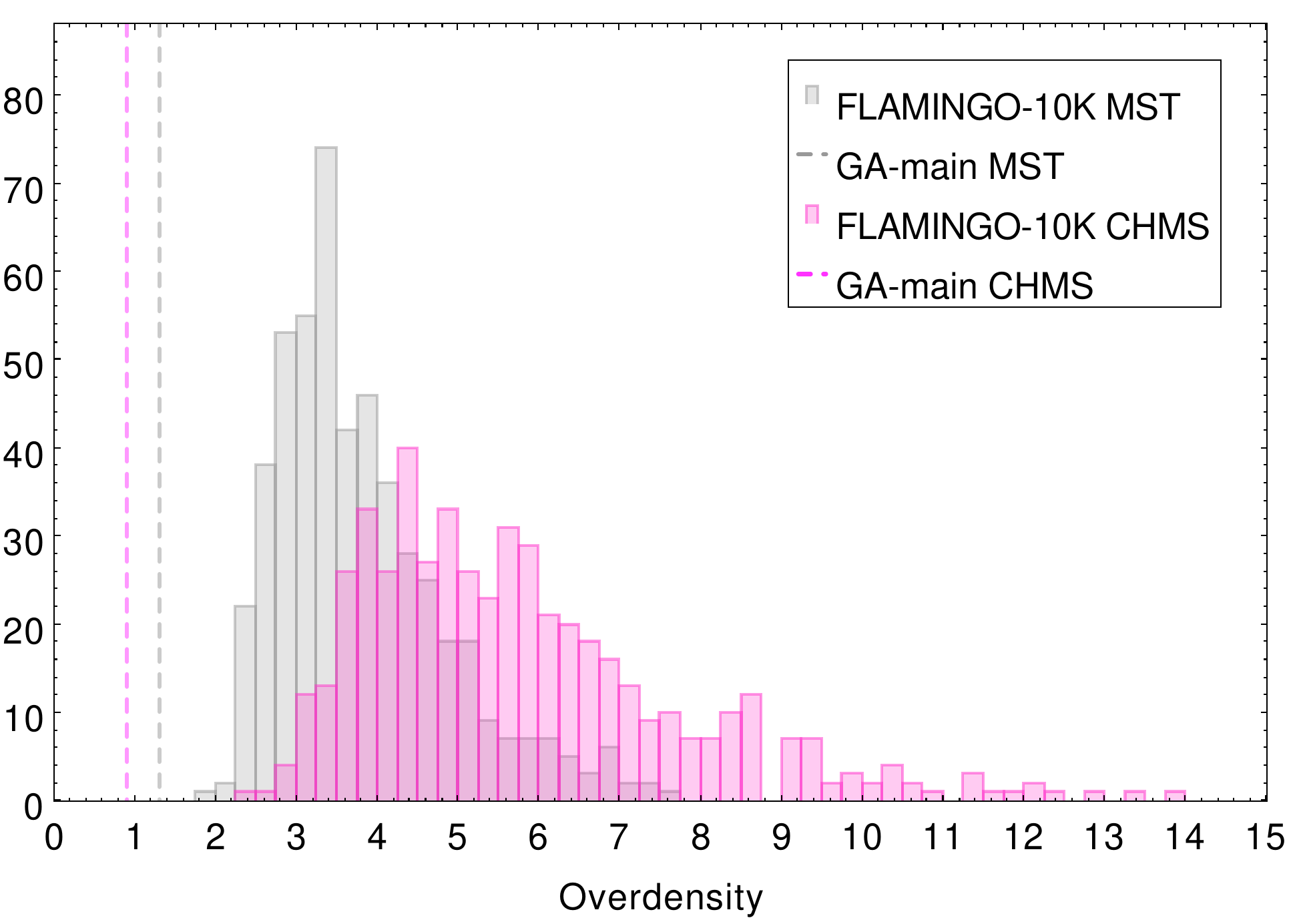}
    \caption{Histogram of the MST and CHMS overdensities (in grey and pink respectively) in $100$ cubes ($2800^3$ Mpc$^3$) of randomly-selected subhaloes in the FLAMINGO-10K simulation data. We applied a linkage scale of $65$~Mpc in each of the $100$ cubes, which is an appropriate choice for simulated data at this field density (see main text). We specified membership $N \ge 10$. Since the candidate structures here are small and clustered, having memberships from the specified minimum of $10$ to a maximum of $19$, the overdensities are correspondingly very high. The histograms have different profiles and are not aligned, demonstrating the difference in the volume estimates with the two methods. Note, however, that these very overdense structures do not have significances, memberships, or sizes that exceed that of GA-main. The GA-main MST and CHMS overdensities are labelled at $1.3$ and $0.9$ with grey and pink vertical, dashed lines, respectively.}
    \label{fig:CHMS_MST_overdens_65Mpc}
\end{figure}

\begin{figure}
   \centering
   \includegraphics[width=0.8\linewidth]{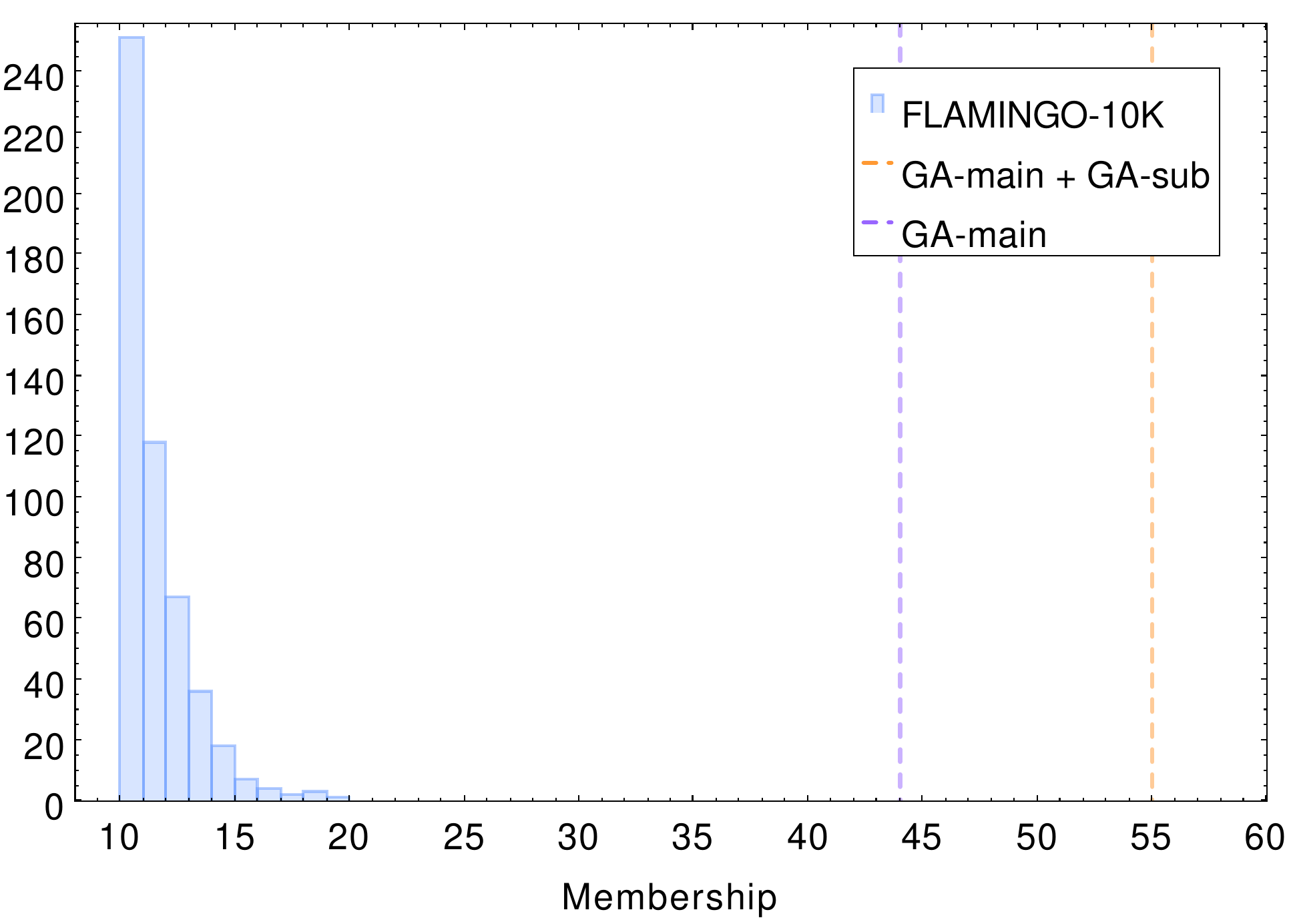}
   \caption{Histogram of the candidate structure memberships (blue) in $100$ cubes ($2800^3$ Mpc$^3$) of randomly-selected subhaloes in the FLAMINGO-10K simulation data. We applied a linkage scale of $65$~Mpc in each of the $100$ cubes, which is an appropriate choice for simulated data at this field density (see main text). We specified membership $N \ge 10$; the largest candidate (by membership) has only $19$ members. The (full) GA (GA-main and GA-sub), and GA-main are labelled at $55$ and $44$ with orange and purple vertical, dashed lines, respectively.}
   \label{fig:membership}
\end{figure}

\begin{figure}
   \centering
   \includegraphics[width=0.8\linewidth]{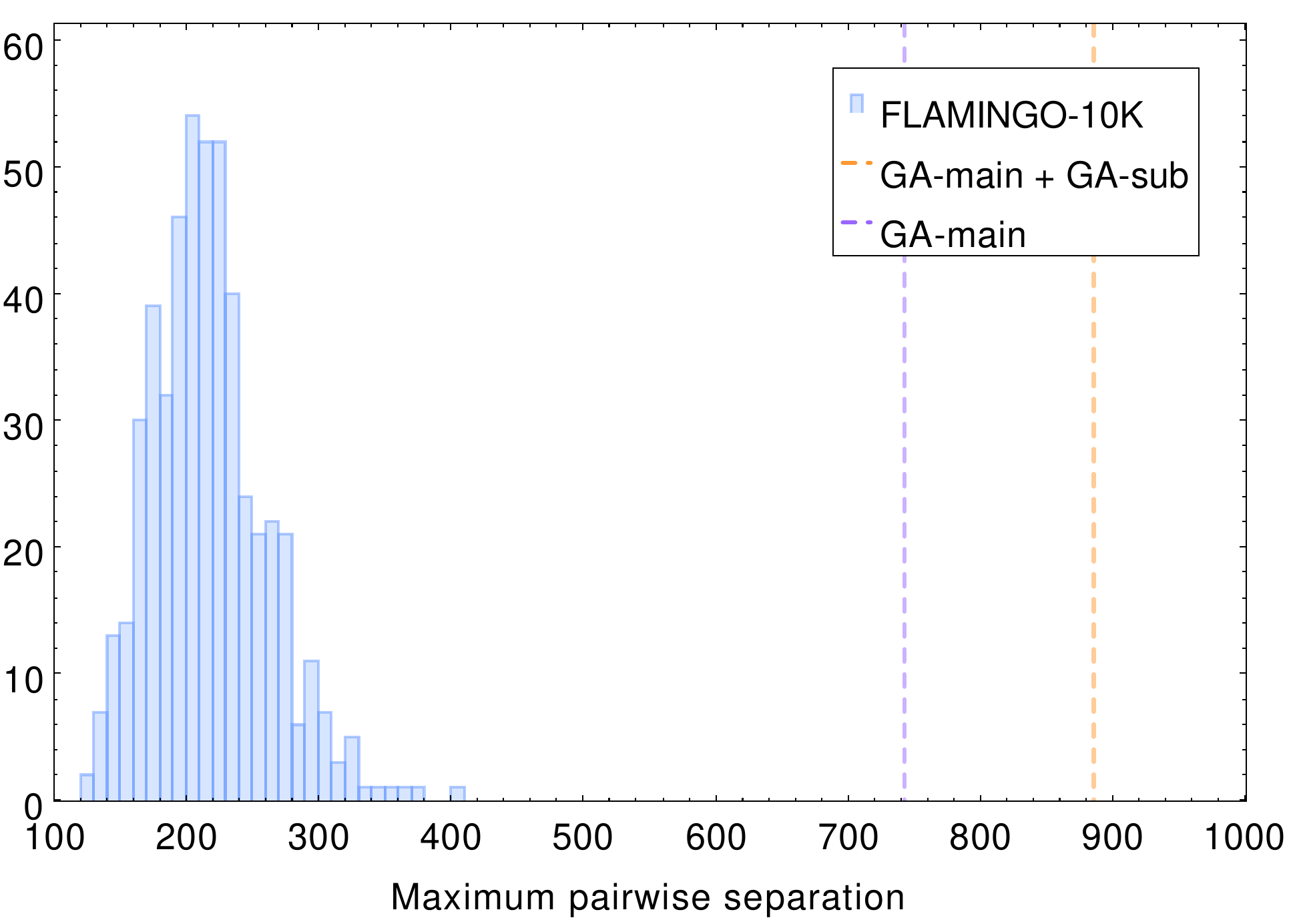}
   \caption{Histogram of the candidate structure maximum pairwise separation (blue) in $100$ cubes ($2800^3$ Mpc$^3$) of randomly-selected subhaloes in the FLAMINGO-10K simulation data. We applied a linkage scale of $65$~Mpc in each of the $100$ cubes, which is an appropriate choice for simulated data at this field density (see main text). We specified membership $N \ge 10$.
   There are no candidate structures identified in the FLAMINGO-10K data that approach or exceed the maximum pairwise separation of the GA (both GA-main and GA-all). The (full) GA (GA-main and GA-sub), and GA-main are labelled at $\sim 886$~Mpc and $\sim 742$~Mpc with orange and purple vertical, dashed lines, respectively.}
   \label{fig:max_pairwise}
\end{figure}

In the GA original discovery paper, we apply the SLHC/CHMS statistical tests to a large field in the SDSS footprint containing the GA (Figure~\ref{fig:GA_big_FOV_in_SDSS}).
The data requested from us and used by Sawala et al. was of the dimensions of the tangent-plane box corresponding to the smaller GA FOV in Figure~$7$ of the GA paper, and the number of Mg~{\sc II} absorbers in this field. 
From these values, they calculated the field density of the smaller GA field and used this as a reference for their random selections in the simulated data. 
As described, the tangent-plane box dimensions of the Mg~{\sc II} in the small GA field requested by Sawala et al. are not the same as the RA-Dec-$z$ box of the large GA field investigated originally with our SLHC/CHMS/MST methods. 
However, the overall field density of Mg~{\sc II} absorbers in the tangent-plane box of the smaller field is equal to that of the field density in the RA-Dec-$z$ box in the larger field, 
so we may continue with the linkage scales calculated earlier.

In Figure~\ref{fig:CHMS_MST_signif_65Mpc} we show the histogram of MST and CHMS significances in $100$ cubes ($2800^3$ Mpc$^3$) of randomly-selected subhaloes in the FLAMINGO-10K data after applying the SLHC algorithm with a linkage scale of $65$~Mpc. We specified membership $N \ge 10$.
The added, dashed, vertical lines show the corresponding MST and CHMS values of GA-main at $-4.7 \sigma$ and $-4.5 \sigma$, respectively. (Our convention is that significances are negative for overdensities.)
The two histograms have similar profiles and are closely aligned, indicating a general agreement between the two different methods (CHMS and MST).
Note that these parameters include all of the structures with $N \ge 10$ identified by the SLHC algorithm, not just those that would exceed the GA in membership and extent (although, there are not any that do).

Figure~\ref{fig:CHMS_MST_overdens_65Mpc} similarly
shows the CHMS and MST overdensities of candidate structures identified in the FLAMINGO-10K data.
The specified minimum membership, $N \geq 10$, is the same as for the GA analysis. 
The largest (by membership) candidate structure has only $19$ members (Figure~\ref{fig:membership}), 
so here we are generally looking at only small, clustered structures.
As would be expected for these small, clustered structures, the MST and CHMS overdensities 
exceed the values of GA-main (represented by dashed, vertical lines).
The histograms showing the CHMS and MST overdensities have different profiles and are not aligned, demonstrating the difference in the volume estimates with the two methods.
Of course, the size, membership and significance of any one of these candidate structures does not exceed that of GA-main or GA-all (GA-main plus GA-sub).
The results highlight, in particular, that a measure of overdensity of a candidate structure is not sufficient on its own; it has to be considered in the context of size and membership, which is an important point missed by Sawala et al. when they compare their `GA-analogues' with the real GA (-main).

In Figure~\ref{fig:max_pairwise} we show the maximum pairwise separation of the candidate structures identified in the FLAMINGO-10K data (blue), with the GA-all and GA-main labelled at $\sim 886$~Mpc and $\sim 742$~Mpc in orange and purple dashed, vertical lines, respectively.
Note that, the pairwise separation is not necessarily a reliable measure of the extent of a structure, should it have intricate morphology such as a spiral, ring or arc shape, for example. However, given that caveat but nevertheless here doing what Sawala et al. do, 
it is clear from the histogram that there are no gigaparsec structures in the FLAMINGO-10K simulation data.
The GA-all and GA-main are significant departures from expectations with FLAMINGO-10K data. 
Thus, here and above, it would seem that the GA is not obviously compatible with the $\Lambda$CDM model.

In our reanalysis of the FLAMINGO-10K simulation data we enhance the previous LSS statistical assessments with the addition of independent reality checks via the CHMS and MST significance tests.
We find no gigaparsec structures, and very few ultra-large large-scale structures (maximum pairwise separation $\geq 370$~Mpc --- that is, above the Yadav-estimated scale of homogeneity \cite{Yadav2010}).
No candidate structures had significance, membership or maximum pairwise separation that approached or exceeded the same parameters in the GA (both GA-main and GA-all).
It might now appear that the GA is a statistically-significant departure in $\Lambda$CDM and presents a more direct challenge.
Conceivably, however, there could be developments in $\Lambda$CDM simulations that would lead to closer approaches to the observed LSS.

\subsection{Comparisons with Poisson data}
\begin{figure}
   \centering
   \includegraphics[width=0.8\linewidth]{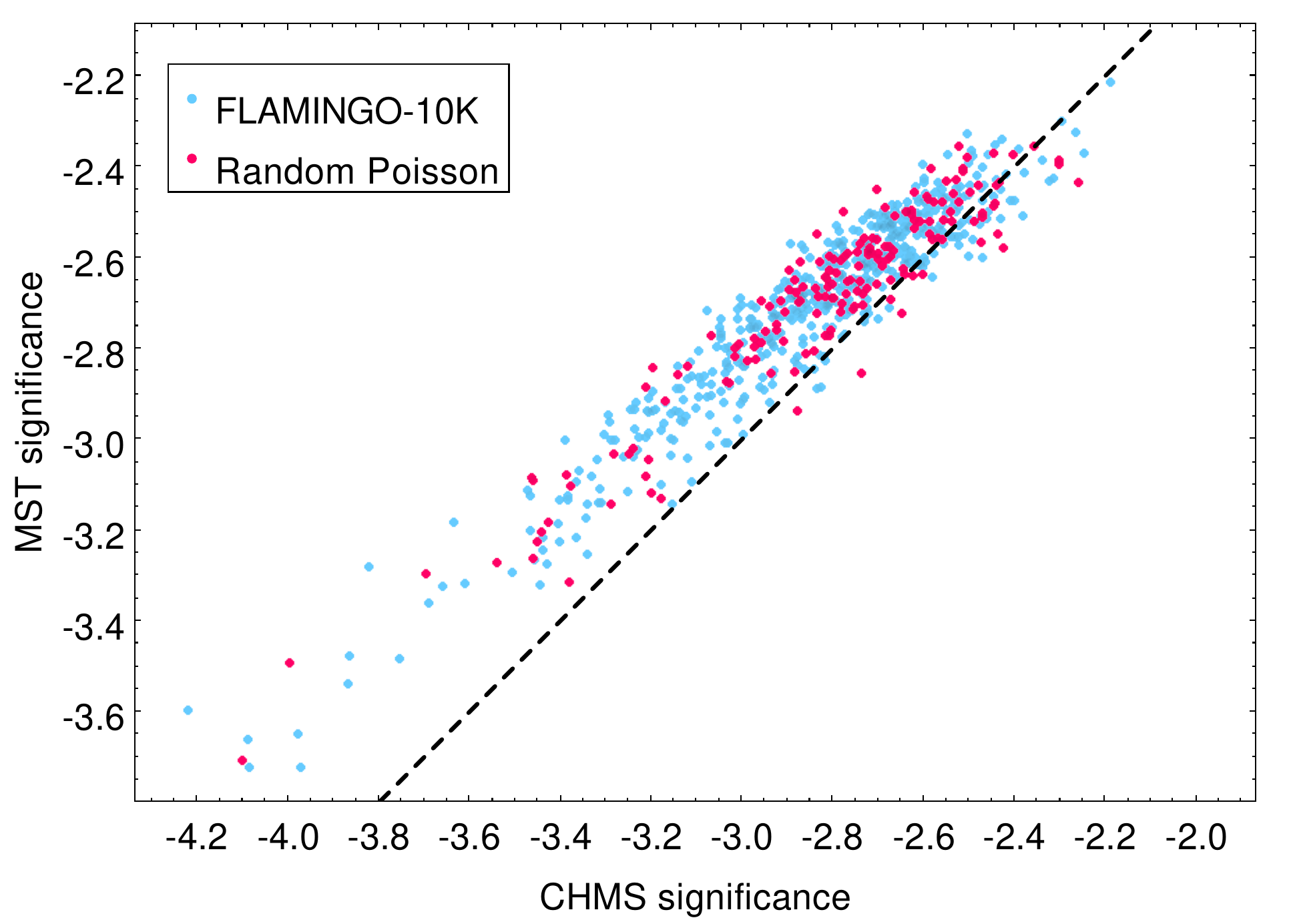}
   \caption{Plot of the MST significance against the CHMS significance for candidate structures identified in the FLAMINGO-10K data (blue) and random Poisson data (magenta). We applied a linkage scale of $65$~Mpc in each of the $100$ cubes ($2800^3$ Mpc$^3$) for the FLAMINGO-10K data and in each of the $40$ cubes ($2800^3$ Mpc$^3$)  for the random Poisson data. We specified membership $N \ge 10$. Both the FLAMINGO-10K and random Poisson data follow the same strong, positive, linear correlation, which suggests that the FLAMINGO-10K can be adequately represented by random Poisson data. In general, the MST significance is slightly smaller (a larger negative number is a higher significance of overdensity) in both datasets (data points above the black, dashed line at $y=x$) than the corresponding CHMS significance.
   }
   \label{fig:CHMS_vs_MST_poisson}
\end{figure}

\begin{figure}
   \centering
   \includegraphics[width=0.8\linewidth]{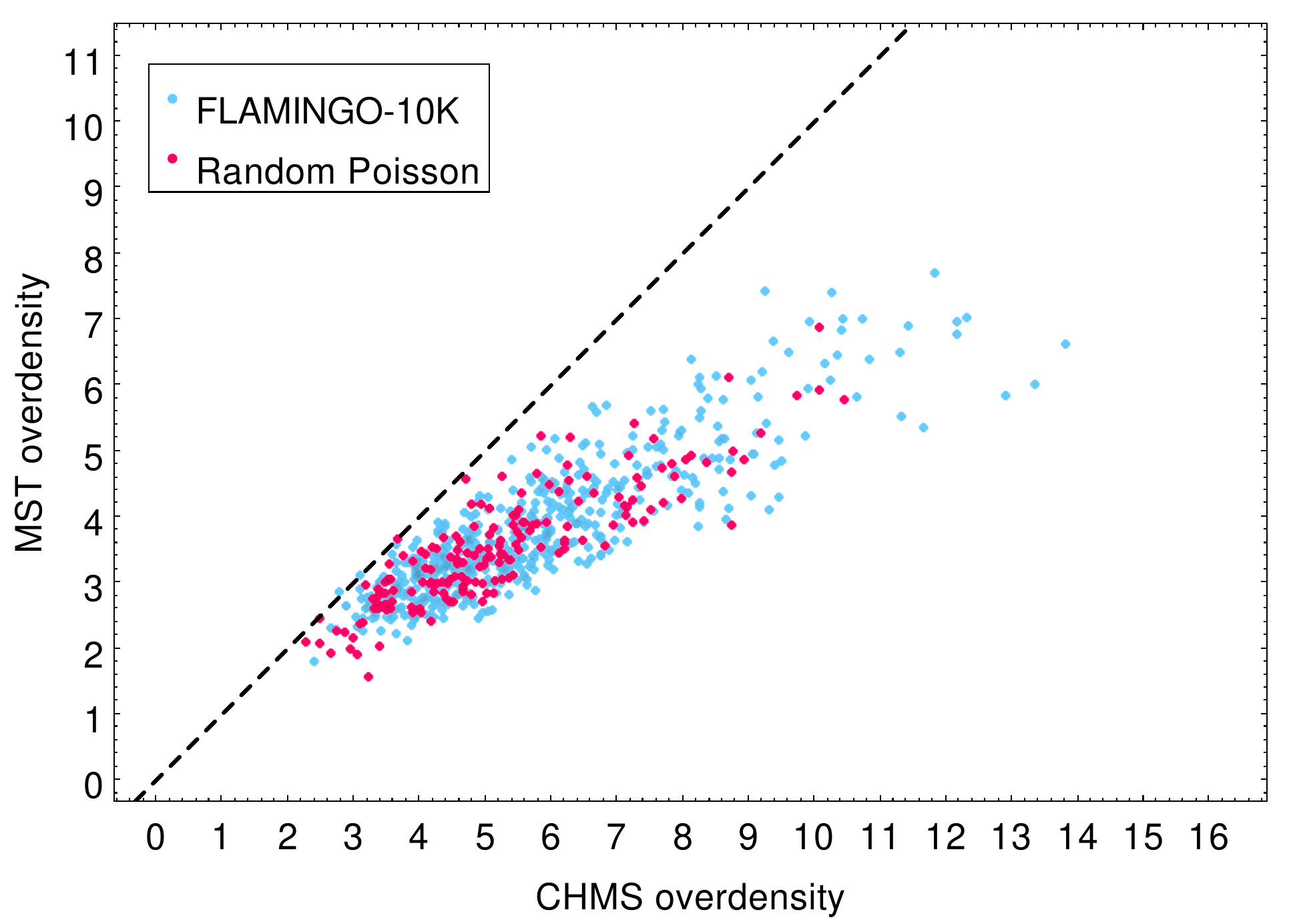}
   \caption{Plot of the MST overdensity against the CHMS overdensity for candidate structures identified in the FLAMINGO-10K data (blue) and random Poisson data (magenta). We applied a linkage scale of $65$~Mpc in each of the $100$ cubes ($2800^3$ Mpc$^3$)  for the FLAMINGO-10K data and in each of the $40$ cubes ($2800^3$ Mpc$^3$)  for the random Poisson data. We specified membership $N \ge 10$. Both the FLAMINGO-10K and random Poisson data follow the same strong, positive, linear correlation, which suggests that the FLAMINGO-10K can be adequately represented by random Poisson data. In general, the MST overdensity is lower in both datasets (data points below the black, dashed line at $y=x$) than the corresponding CHMS overdensity. Given that the candidate structures are likely small, clustered groups (given the small memberships), we might expect CHMS to measure a higher overdensity, as the CHMS is known to favour small, clumpy structures. 
   }
   \label{fig:CHMS_vs_MST_overdens_poisson}
\end{figure}

\begin{figure}
   \centering
   \includegraphics[width=0.8\linewidth]{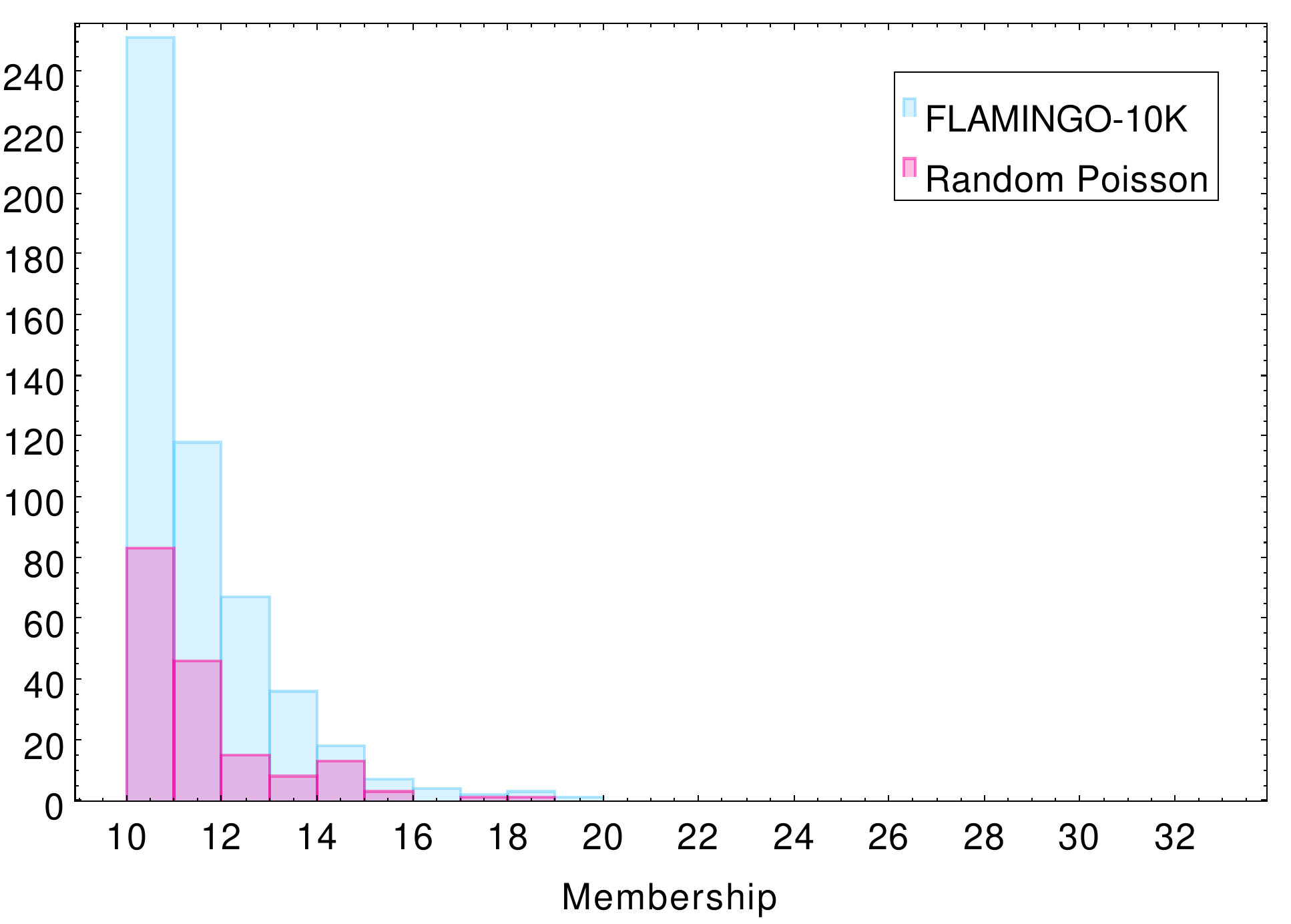}
   \caption{Histogram of the memberships for candidate structures identified in the FLAMINGO-10K data (blue) and random Poisson data (pink). We applied a linkage scale of $65$~Mpc in each of the $100$ cubes ($2800^3$ Mpc$^3$)  for the FLAMINGO-10K data and in each of the $40$ cubes ($2800^3$ Mpc$^3$)  for the random Poisson data. We specified membership $N \ge 10$. The two histograms are aligned, having extremely similar profiles, suggesting that the FLAMINGO-10K data can be adequately represented by random Poisson data. 
   }
   \label{fig:membership_poisson}
\end{figure}

\begin{figure}
   \centering
   \includegraphics[width=0.8\linewidth]{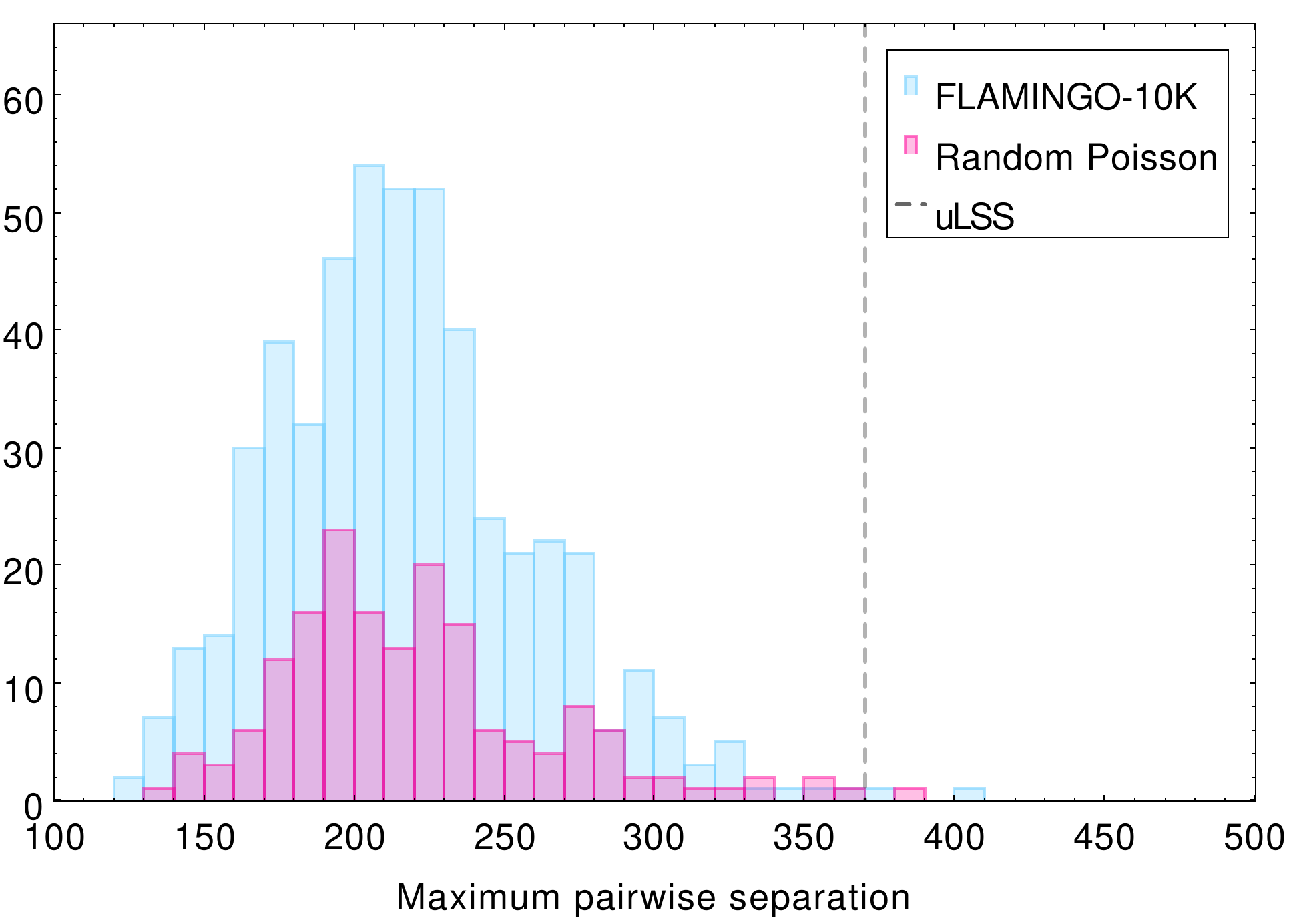}
   \caption{Histogram of the maximum pairwise separation for candidate structures identified in the FLAMINGO-10K data (blue) and random Poisson data (magenta). We applied a linkage scale of $65$~Mpc in each of the $100$ cubes ($2800^3$ Mpc$^3$)  for the FLAMINGO-10K data and in each of the $40$ cubes ($2800^3$ Mpc$^3$) for the random Poisson data. We specified membership $N \ge 10$. The two histograms are closely aligned, having extremely similar profiles, suggesting that the FLAMINGO-10K data can be very closely approximated by random Poisson data. The random data appears to peak at a slightly lower separation value, suggesting that the FLAMINGO-10K data detects slightly larger structures. The added dashed, grey, vertical line is where we have previously defined `ultra-large' large-scale structures (uLSSs) at $\sim 370$~Mpc. A small fraction of the candidate structures in both FLAMINGO-10K and random Poisson exceed the uLSS limit, but there are no gigaparsec structures found at all.
   }
   \label{fig:max_pairwise_poisson}
\end{figure}

For a final check, we show what arises when our enhanced LSS statistical methods are applied to random Poisson point data. 
We use the subhalo selection from the data provided by Sawala, for masses in the range $1$--$5 \times 10^{12}$M$_{\odot}$, as previously.
We then replace the coordinates of the Sawala points with random coordinates between $(0, 2800)$~Mpc.
In general, we find that the FLAMINGO-10K data can be closely approximated by random Poisson data (figures~\ref{fig:CHMS_vs_MST_poisson} to \ref{fig:max_pairwise_poisson}).  

In Figure~\ref{fig:CHMS_vs_MST_poisson} we are seeing the spread of MST significance against CHMS significance in candidate structures identified in the FLAMINGO-10K data (light blue) and random Poisson data (magenta) with a linkage scale of $65$~Mpc (an appropriate linkage scale for these data). 
The trends in the two datasets are very similar: they both are strongly, positively, linearly correlated, favouring slightly higher CHMS significances over MST significances ($y=x$ is shown as the black, dashed line).
Additionally, the typical candidate structures identified in the Poisson data are small, having low memberships (Figure~\ref{fig:membership_poisson}), and rarely exceeding a maximum pairwise separation of $370$~Mpc (for uLSSs; Figure~\ref{fig:max_pairwise_poisson}), which we saw was also the case for the FLAMINGO-10K data.

The candidate structures identified in both datasets (FLAMINGO-10K and Poisson) appear likely to be small, clumpy-like structures, where the CHMS can easily define a volume encompassing all of the members without incorporating empty volumes (a problem that would arise in filamentary structure).
Perhaps what is surprising is the lack of longer, filamentary-type structures seen in the FLAMINGO-10K data compared with the Poisson data.
We have seen in real data that long, filamentary walls and structures are found, such as the GA and BR in Mg~{\sc II} absorbers, and others such as the Sloan Great Wall \cite{Gott2005}, the South Pole Wall \cite{Pomarede2020}, and the recently-discovered Quipu structure \cite{Bohringer2025}, in other types of surveys.
However, many structures, such as superclusters and walls, are rarely analysed with independent reality-assessment tests (perhaps they are \emph{obvious}), which makes them hard to compare.

In Figure~\ref{fig:CHMS_vs_MST_overdens_poisson} we are seeing the spread of MST overdensity against CHMS overdensity in candidate structures identified in the FLAMINGO-10K data (light blue) and random Poisson data (magenta) with a linkage scale of $65$~Mpc (an appropriate linkage scale for these data). 
The trends in the two datasets are very similar: they both are strongly, positively, linearly correlated, favouring slightly higher CHMS overdensity over MST overdensity ($y=x$ is shown as the black, dashed line); this aligns well with what we saw previously with the significances.
Given that we are likely looking at small, clumpy-like structures, it is unsurprising to find that the structures are very overdense.
Of course, their sizes, memberships and significances do not approach or exceed the GA values.

The results in this section show that the FLAMINGO-10K data can be adequately represented by random Poisson data. 
Perhaps it is a little concerning that there is not a larger discrepancy between what is found in random data, compared with what is found with advanced cosmological simulations.

\section{Discussion and conclusions} 
In this paper we had two objectives: (i) to address the shortcomings of the analysis presented by Sawala et al. 2025 \cite{Sawala2025} and (ii) to reanalyse the LSS in FLAMINGO-10K simulations.
We have explored the several shortcomings in the Sawala et al. analysis, and we hope to dispel the misconception that finding patterns in simulations is noteworthy or the equivalent of detecting and analysing real cosmological structure.
A summary of the defects in the Sawala et al. analysis is as follows.

(1) The `GA-analogues' in Sawala et al. did not in fact exceed both the GA membership and overdensity together (even with their large `look-elsewhere' effect). Consider a small cluster of points: the volume around these points will be small and the cluster will likely have a high density. However, this cluster would not necessarily (and usually not at all) be considered a significant deviation from random Poisson expectations. Finding a structure with high membership, high overdensity, and a significant deviation from random expectations is noteworthy. Sawala et al. do not find evidence for such structures in their analysis.

(2) When applying their FoF algorithm, Sawala et al. use an inappropriate linkage scale of $95$~Mpc. They fail to recognise that the linkage scale that is appropriate for real data has to incorporate allowances for redshift errors, peculiar velocities, and observational artefacts. For simulated data, the redshifts and positions of the data points are known exactly, so an appropriate linkage scale for this data would follow from the mean nearest-neighbour separation: $\bar{r} \approx 0.55 (1/\rho)^{1/3}$, which is $65$~Mpc in their case. 
In applying an inappropriately-large linkage scale to the FLAMINGO-10K data, Sawala are favouring the undesirable effect of chaining \cite{MurtaghHeck}.

(3) The authors do not apply an independent reality-assessment test for analysing the statistical significance of candidate structures or `GA-analogues'. Without such a test, patterns detected in noise might be mistakenly classed as cosmologically-interesting structures. 

(4) The simulated data are acquired at one specific redshift, whereas real data are unavoidably measured at continuous redshifts. Therefore, when choosing to look at small / thin redshift slices in real data, we are attempting to avoid the complications of redshift evolution. In simulated data, redshift slices need not be a limitation to investigating LSS.

(5) Sawala et al. investigate the structure persistence of the GA-analogues. They find no evidence for structure persistence which indicates that their patterns are not tracing any real underlying matter. They then illogically conclude that the GA must also not trace any real underlying matter. Their conclusion is a non sequitur; they do not, in fact, show that the GA is not tracing any real underlying structure. Furthermore, given that we have been unable to detect structures with either MST or CHMS significance equal to or exceeding the GA-main in our reanalysis of the FLAMINGO-10K data, we suspect that the structures detected in the simulations are more likely attributable to noise, whereas the real GA is a significant structure that is most likely tracing real underlying matter.

(6) In the GA original discovery paper, we not only applied three different statistical tests (only one of which is addressed by Sawala et al.), but we investigated the observational properties of the GA. The GA was supported by independent corroboration of field quasars, and was also found to have an interesting redshift distribution along the arc, which appeared somewhat reflected with the equivalent width distribution of the Mg~{\sc II} lines. Subsequently, we made a further discovery, of the Big Ring, just $\sim 12^{\circ}$ north of the GA.
The further discovery of another remarkable uLSS in the same cosmological neighbourhood as the GA must surely be rare. Finding one such remarkable structure in FLAMINGO-10K simulations proved to be difficult (impossible); what about finding two such structures at the same location?  
We have previously noted that the \emph{accumulating set} of uLSS discoveries (of all types) could be important for cosmology.
Presumably, 
following where these discoveries might lead could further expand our knowledge and understanding in cosmology, or even hint towards new developments beyond $\Lambda$CDM.

We would like to emphasise that the SLHC algorithm, or other MST-type tests, are not strictly applicable or appropriate to the Mg~{\sc II} data, given the complications of the data.
Unfortunately, as the GA and BR analyses were necessarily post-hoc, we used these statistical tests (and others) to tell us more about the properties of the LSSs and their connectivity. 
Ideally, we should try to develop a more appropriate statistical test.
However, in the meantime, one \emph{must} proceed with caution when using the MST-type algorithms on data with intrinsic spatial variations. 

In our reanalysis of the FLAMINGO-10K subhaloes we have shown that gigaparsec structures are nowhere to be seen.
Now that we have investigated more appropriately the statistical LSS properties of the FLAMINGO-10K subhaloes and compared the outcomes with the Mg~{\sc II} data, perhaps we can speculate that an answer to the discrepancies will eventually be attributed to a mismatch on large scales of $\Lambda$CDM (simulations) with the real Universe, or, perhaps more simply, to algorithmic artefacts in the simulations \cite{Moreno2025}.
Alternatively, there are, of course, those who might conclude that the results presented here represent a direct challenge to $\Lambda$CDM.

\acknowledgments
AML thanks the University of Lancashire (formerly Central Lancashire) for postdoctoral funding.
We acknowledge the use of the public R software (v4.1.2)\footnote{https://www.R-project.org/}.  Our data has depended on the
publicly-available Sloan Digital Sky Survey quasar catalogue and the
corresponding Mg~{\sc II}
catalogues from Zhu and M\'enard 2013\footnote{https://www.guangtunbenzhu.com/jhu-sdss-metal-absorber-catalog}. 
We thank Till Sawala for providing us with the FLAMINGO-10K subhalo data that they used in their analysis.

\end{document}